\documentclass[epj,nopacs,final]{svjour}
\usepackage{latexsym}
\usepackage{graphicx}
\usepackage{epsfig}

\usepackage{amsmath}
\newcommand{\be}{\begin{equation}}
\newcommand{\ee}{\end{equation}}

\begin{document}
\title{Light-nuclei production and search for the QCD critical point}
\author{Edward Shuryak\inst{1} \and Juan M. Torres-Rincon\inst{2}}                    

\mail{torres-rincon@itp.uni-frankfurt.de}
\institute{Department of Physics and Astronomy, Stony Brook University,
Stony Brook, NY 11794-3800, USA \and Institut f\"ur Theoretische Physik, Johann Wolfgang Goethe-Universit\"at, Max-von-Laue-Strasse 1, D-60438 Frankfurt am Main, Germany}
\date{Received: 8 May 2020 / Accepted: 7 September 2020}
%
\abstract{
We discuss the potential of light-nuclei measurements in heavy-ion collisions at intermediate energies for the search of the hypothetical QCD critical end-point. A previous proposal based on neutron density fluctuations has brought appealing experimental evidences of a maximum in the ratio of the number of tritons times protons, divided over deuterons square, ${\cal O}_{tpd}$. However these results are difficult to reconcile with the state-of-the-art statistical thermal model predictions. Based on the idea that the QCD critical point can lead to a substantial attraction among nucleons, we propose new light-nuclei multiplicity ratios involving $^4$He in which the maximum would be more noticeable. We argue that the experimental extraction is feasible by presenting these ratios formed from actual measurements of total and differential yields at low and high collision energies from FOPI and ALICE experiments, respectively. We also illustrate the possible behavior of these ratios at intermediate energies applying a semiclassical method based on flucton paths using the preliminary NA49 and STAR data for ${\cal O}_{tpd}$ as input.
%
} 
\maketitle
\section{Introduction}
\label{intro}

Significant efforts, in both theory and experiment, have been devoted in heavy-ion collision physics to the search of the hypothetical QCD critical point and first-order transition between confined and deconfined matter~\cite{Luo:2017faz}. For the former, the bulk of these studies follows a common driving idea~\cite{Stephanov:1998dy,Stephanov:1999zu}: the search of indicative observables showing a nonmonotonous behavior when measured as functions of a control parameter (collision energy, system size, rapidity acceptance...). The ultimate reason is the existence of physical quantities which present critical enhancement with the correlation length close to $T_c$~\cite{Luo:2017faz,Stephanov:2008qz}. Such critical behavior should manifest as a nonmonotonous behavior when approaching and subsequently overpassing the critical end-point.

An example of such observables have been the higher-order cumulants of the net-baryon distribution~\cite{Stephanov:2011pb}. In particular, the scaled kurtosis of the net-proton distribution as a function of the collision energy has led to some clues that might be related to the critical dynamics~\cite{Stephanov:2011pb,Luo:2015ewa,Adam:2020unf}. In Ref.~\cite{Shuryak:2018lgd} we focused on nucleon dynamics and the $NN$ interaction as a contributing source for these quantities. The attractive part of the pairwise potential, as being dominated by the long-ranged critical ($\sigma$) mode of QCD~\cite{Serot:1984ey}, is expected to have an important role in these observables.

Focusing on more global observables we proposed that the increase of the nuclear attraction might lead to an extra production of light nuclei if the critical mode acts substantially among nucleons. In a favorable situation, we expect an enhanced production of light nuclei close to $T_c$ with respect to a noncritical scenario~\cite{Shuryak:2019ikv}.

The statistical thermal model (STM)~\cite{Andronic:2017pug} has been used as a good tool to describe multiplicities of hadronic states $N_i$ at high energies ($i$ denotes the type of hadron). In particular, its application to hadrons and light nuclei has been successfully tested at LHC energies~\cite{Braun-Munzinger:2018hat}. The application of this model typically ignores the effects of the $NN$ potential (although extensions of it do include interactions between baryons e.g.~\cite{Poberezhnyuk:2019pxs}). A sizable modification of the $NN$ potential would bring a small positive correction to the light-nuclei multiplicities $\delta N_i$. One expects that such modification cannot be observed when looking at total nuclear yield, as $\delta N_i \ll N_i$. However one can construct certain nuclear multiplicity ratios in which the common thermodynamic factors (temperature, chemical potentials, volume...) cancel out for thermal abundances.  Only the presence of nonideal thermodynamic effects (e.g. when the interaction potential is comparable to the temperature) makes these ratios non trivially dependent on the collision energy, giving a more sensitive observable to detect such modifications~\cite{Shuryak:2019ikv}. 

STMs implement feed down corrections from higher unstable hadronic states and excited nuclei~\cite{Hahn:1986mb}. This is a relevant effect which must be included in the calculation of these nuclear ratios. It turns out~\cite{Vovchenko:2020dmv} that after the implementation of these feed down effects, some these ratios might remain, after all, rather flat in the relevant collision energy region. Therefore, a further attractive $NN$ potential---comparable to the typical freeze-out temperatures---would still introduce corrections producing eventual nonmonotonous ratios in consonance with the location of the critical region~\cite{Shuryak:2018lgd}.

In the following we will review the recent theoretical advances in this direction, and motivate the study of such ratios by presenting the status of the ratio ${\cal O}_{tpd}$ involving tritons, protons and deuterons. Later, we will introduce different proposals in which the effect of the critical point is expected to be enhanced, and motivate our experimental colleagues to address these ratios by presenting some experimental results measured away from the critical region, and the expectations coming from a novel theoretical approach based on a semiclassical method at finite temperature. 

\section{"Preclusters" and the special role of four-nucleon systems}

It is a well-known fact that nuclear forces include two delicate cancellations, between the attractive and repulsive parts of the $NN$ potential, as well as between potential and kinetic energies~\cite{Serot:1984ey}. As a result, neutron systems are all unbound, while deuteron and triton have only one shallow state. The champion in small binding is perhaps the $pn\Lambda$ system, with an exact value of the binding energy still in dispute~\cite{Juric:1973zq,Adam:2019phl}.

However the situation drastically changes for the four-nucleon systems. The $^4$He binding energy of 28 MeV is much larger than that of lighter systems. And---as we pointed out in~\cite{Shuryak:2019ikv}--- unlike these, it possesses multiple bound and resonant states. Their decay should feed down into production of lighter systems. Even if they eventually decay back into four independent nucleons, their correlation in phase space is important, contributing positively to the kurtosis of the proton multiplicity distribution~\cite{Shuryak:2018lgd}.

In our previous work~\cite{Shuryak:2019ikv} we introduced the notion of ``preclusters'', or statistical spatial correlation of several nucleons at the kinetic freeze-out. While matter reaches a good degree of equilibration by that time, and formation of preclusters themselves is subject to equilibrium statistical mechanics, the subsequent transition of preclusters to  clusters---light nuclei states or resonances at zero temperature and density---have a completely different temporal and spatial scales and is dynamical.  For example, for four nucleons (to be mostly discussed) the splitting between states and widths all have the magnitude of few MeV, and thus their separation and decay require times of hundreds of fm$/c$. While preclusters are compact objects, with sizes $1-2$ fm, the clusters and their decay products are much larger in size. This distinction helps to understand a well-known paradox of apparently copious production of large-size and fragile clusters. 

Quantum statistical calculations of four nucleons is in practice not trivial, as the system has $9$ coordinates. In~\cite{Shuryak:2019ikv} we also developed two theoretical tools able to calculate properties and production rates of preclusters. We extended a novel semiclassical method for the density matrix, based on special classical paths called ``fluctons''~\cite{Shuryak:1987tr,Escobar-Ruiz:2016aqv} to finite temperatures. We now prepare a separate methodical paper~\cite{thermal_flucton} on this method, as its applications can be extended well beyond heavy-ion physics.

Alternatively, we extended the method of hyperspherical harmonics, previously used only for ground states of few-nucleon systems, to calculation of ${\cal O}(100)$ quantum states, which also allowed to calculate the thermal density matrix. 

Furthermore, in a soon-coming work~\cite{DeMartini} the most straightforward quantum mechanical method, that of path integral simulations, is used. It allows to calculate quantum effects for many-body systems in multidimensional settings, without any approximations. 

In all of the above mentioned works the ``preclustering'' phenomenon is well documented, both for the standard (unmodified) nuclear forces and those with potential modifications, due to either finite temperature-density effects or a hypothetical critical point. In all of them one observes the same basic
fact: preclustering is extremely sensitive to forces, especially at large distances. Therefore, light nuclei production must be
a good observable to study.

Later we will apply the finite-temperature flucton method described in Ref.~\cite{Shuryak:2019ikv} to obtain the distribution of the relative nucleon distance as a function of the $NN$ potential. For the latter, we will make use of the Serot-Walecka potential as given in~\cite{Shuryak:2018lgd,Shuryak:2019ikv}, whose attractive part is dominated---close to the critical point---by the $\sigma$-excitation mass. This method will be used to compute several nuclear multiplicity ratios. Before, we start by reviewing the experimental situation of one of these special ratios.

\section{Triton-proton-deuteron ratio}
\label{sec:1}

Based on the coalescence model for nuclei production, the authors of Ref.~\cite{Sun:2017xrx} considered neutron density fluctuations quantified by $\Delta n=\langle (\delta n)^2\rangle/\langle n\rangle^2$, which is sensitive to the QCD critical point. This quantity controls the ratio
\be \label{eq:ratio1coa} {\cal O}_{tpd} \equiv \frac{N_t N_p}{N_d^2} \simeq 0.29 (1+\Delta n) \ , \ee
which involves tritons, protons and deuterons. The coefficient $0.29$ is a combination of numerical factors and spin degeneracies of the involved nuclei. In a situation with negligible fluctuations one simply has ${\cal O}_{tpd}=0.29$, independent of the collision energy.

From a different perspective, the pure thermodynamical production of a nucleus $i$ composed by $A$ nucleons is the Boltzmann factor describing the well-known exponential suppression with $A$~\cite{Andronic:2017pug}. The equilibrium number of nuclei is
\be \label{eq:yield} N_i \simeq \frac{g_i {\cal V}}{(2\pi)^{3/2}} (m_i T)^{3/2}  \exp\left(-\frac{m_i}{T}\right) \exp \left(\frac{\sum_{a} Q_i^a \mu_i^a}{T} \right)\ , \ee
where $g_i$ is the spin degeneracy of nucleus $i$, ${\cal V}$ the volume, $m_i \simeq Am_N$ the mass of the nucleus, $T$ the temperature and $Q_i^a,\mu_i^a$ the corresponding charges and chemical potentials ($a=\{B,Q,S \})$. 

By forming multiplicity ratios in which this exponential factor cancels out, one might obtain some interesting results. As we emphasized in~\cite{Shuryak:2019ikv}, that should come from different powers of the interaction potential per nucleon pair. In particular, for the ratio in~(\ref{eq:ratio1coa}),
\be \label{eq:ratio1therm} {\cal O}_{tpd} \simeq 0.29 \frac{\langle e^{-3V(r)/T} \rangle}{\langle e^{-V(r)/T}\rangle^2 } \ , \ee
where the thermal average $\langle \rangle$ reflects the nonideal contribution of the internucleon potential $V(r)$ averaged in space.  In the ideal case~(\ref{eq:yield}), where the interaction energy is negligible with respect to temperature, one again obtains ${\cal O}_{tpd}=0.29$.

While a rigorous treatment of clusterization in the vicinity of the critical point is not yet developed, it is intuitively clear that the increase of attraction might lead to formation of precluster structures, if the nucleons can feel the presence of the critical region long enough. Therefore, the conceptual conclusion in both approaches (\ref{eq:ratio1coa}) and (\ref{eq:ratio1therm}) is similar: close to $T_c$ both $\Delta n$ and $|V(r)/T|$ might become ${\cal O}(1)$ and the ratio will present a maximum for those collisions passing close to the QCD critical point.

The authors of Ref.~\cite{Sun:2017xrx} presented evidences of such a maximum around $\sqrt{s_{NN}}=8.8$ GeV in the results of the NA49 experiment~\cite{Anticic:2016ckv} (a second peak was also suggested in~\cite{Sun:2018jhg}). More recently, preliminary STAR data showed a more prominent peak in ${\cal O}_{tpd}$ around $\sqrt{s_{NN}} =27$ GeV~\cite{Zhang:2020ewj}. These results are really intriguing. However the main puzzle resides in the comparison of the experimental data with the up-to-date calculations using the STM, and also coalescence studies~\cite{Oliinychenko:2020ply}. Not only there is a lack of agreement among models, but also between theory and experimental data. One of the known problems~\cite{Zhang:2020ewj} is the apparent overestimate of tritons by the STM, whereas protons and deuterons seem to be well described~\cite{Vovchenko:2020dmv,Zhang:2020ewj}.

\begin{figure}[ht]
\centering
\resizebox{7.2cm}{!}{\includegraphics{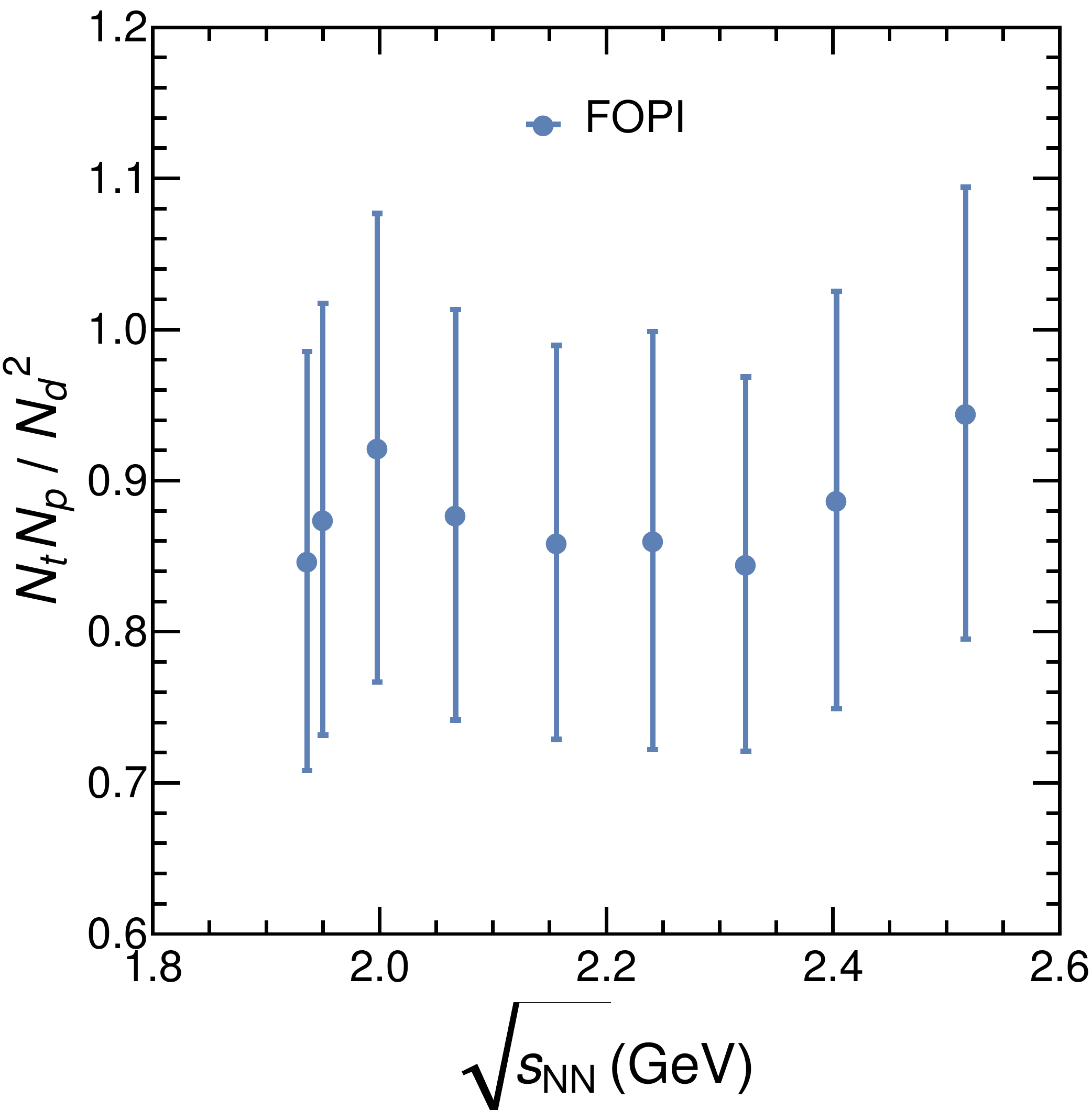}}
\resizebox{7.2cm}{!}{\includegraphics{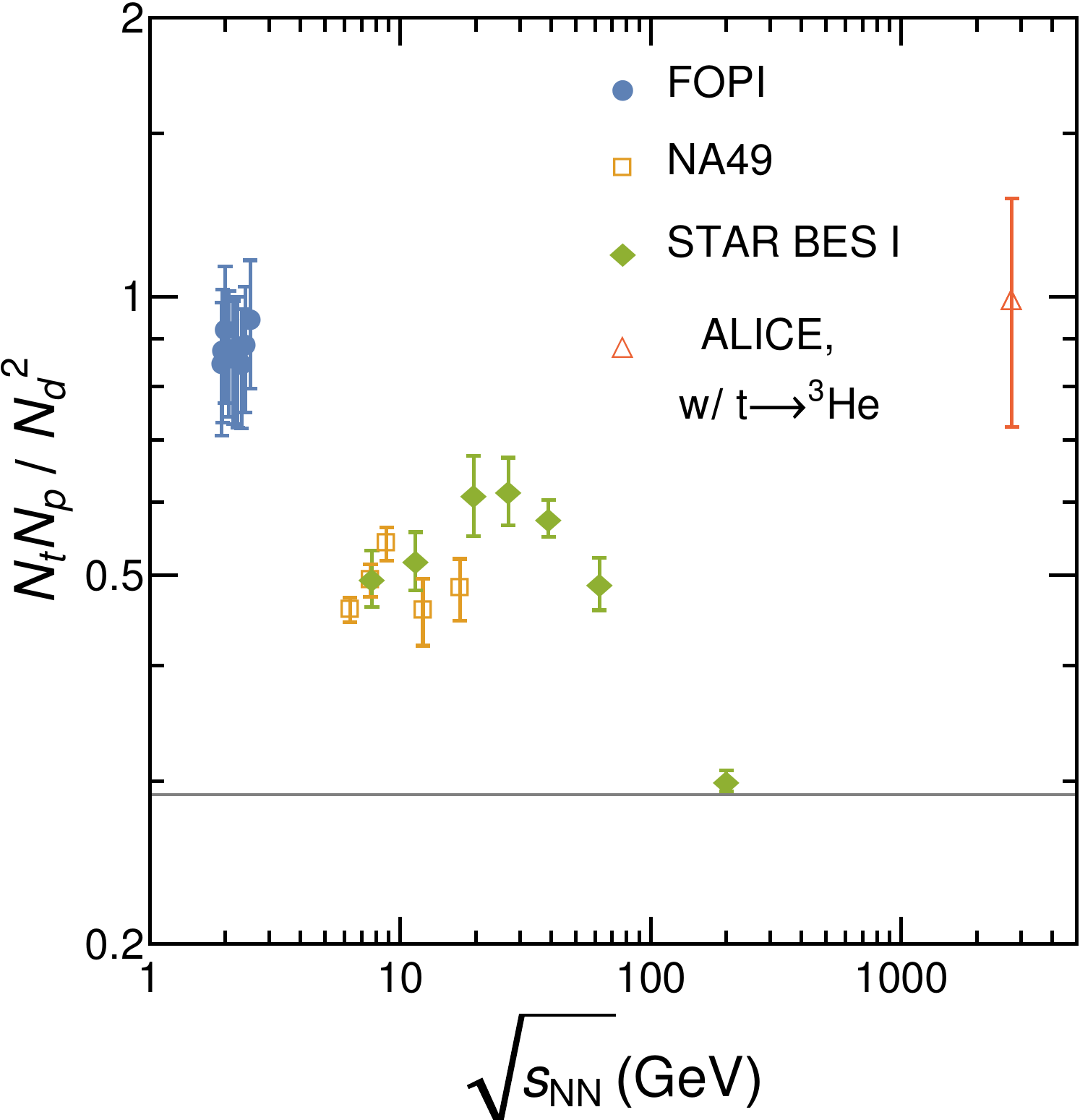}}
\vspace*{0.2cm}
\caption{${\cal O}_{tpd}$ ratio as a function of the center-of-mass collision energy. Upper panel: Ratio formed from FOPI multiplicity results at low energies~\cite{Reisdorf:2010aa}. Bottom panel: Same FOPI-based data together with NA49~\cite{Anticic:2016ckv}, STAR~\cite{Zhang:2020ewj} and ALICE~\cite{Abelev:2013vea,Adam:2015vda,Acharya:2017bso} preliminary results. NA49, STAR and ALICE results are based, not in ratios of total multiplicities, but in $dN/dy$ ratios at midrapidity, and in the later, the triton yield is traded by $^3$He. The horizontal line indicated the value 0.29, cf.~Eq~(\ref{eq:ratio1therm}).}
\label{fig:1}       
\end{figure}

Focusing on experimental results, we present the ratio ${\cal O}_{tpd}$ formed from published FOPI data~\cite{Reisdorf:2010aa} at low energies in the upper panel of Fig~\ref{fig:1}. FOPI Collaboration has collected yields of several nuclei in heavy-ion collisions for 25 combination of colliding heavy-nuclei and collision energies. We focus on the most central Au+Au collisions for the beam energies (per nucleon) $E/A=$0.12, 0.15, 0.25, 0.4, 0.6, 0.8, 1.0, 1.2, 1.5 GeV (we eventually show them converted to center-of-mass energy). From the final products we consider $p, d, t, ^3$He and $^4$He, measured after a $4\pi$ reconstruction~\cite{Reisdorf:2010aa}. From the reported total yields we build the needed light-nuclei multiplicity ratio, and assign an error bar by propagating the experimental uncertainties.

We observe that the ratio ${\cal O}_{tpd}$ using FOPI results is rather flat, and the data might be already affected by the presence of the multifragmentation regime at low energies. In the lower panel, we plot together the FOPI results with NA49 data~\cite{Anticic:2016ckv}, and the preliminary STAR~\cite{Zhang:2020ewj} and ALICE~\cite{Abelev:2013vea,Adam:2015vda,Acharya:2017bso} results. While FOPI uses $4\pi$-reconstructed multiplicities, the other experiments consider $dN/dy$ values at midrapidity, instead of total yields. Moreover tritons in ALICE are difficult to disentangle from other hadrons via energy loss only, and additional analyses (with the time-of-flight detector) are needed. Yet it is possible to substitute their yields by $^3$He, which is expected to be similar at these energies based on isospin symmetry~\footnote{We thank Benjamin D\"onigus for pointing this out to us~\cite{Braun-Munzinger:2018hat}.}.

It is interesting that FOPI data at its top energies are compatible with the ALICE results. However ALICE point is known to be above the simple-minded thermal value for a well-known reason: protons get a significant contribution from feed down of excited baryons~\cite{Andronic:2017pug}. There is no feasible feed down to $t,d$~\cite{Vovchenko:2020dmv}. The opposite situation is at FOPI energies: here temperature is too low to excite baryons, but $\mu_B$ is close to the nucleon mass and there should be large clustering feed downs, both to $t$ and $d$~\cite{Vovchenko:2020dmv}. Therefore, the similar result of FOPI and LHC must be a coincidence. 

It is also curious to observe that while the degree of feed-down should be similar, the STAR point with the smallest error (corresponding to the main energy of RHIC operations, $\sqrt{s_{NN}} =200$ GeV) is away from the ALICE point, and close to the Boltzmann prediction (without feed-down). As we already said, ALICE data was very well described by the STM~\cite{Braun-Munzinger:2018hat}, and the STAR result on triton multiplicity~\cite{Zhang:2020ewj} presents a discrepancy with the STM prediction. This discrepancy remains for lower energies, and should certainly be addressed in the future.

Both NA49 and STAR data tell us that some nontrivial dynamics might be happening at intermediate energies, with an important medium modification with respect to the ideal gas case. Among them, there is the possible peak signaling the critical point, in accordance with the physics encoded in Eqs.~(\ref{eq:ratio1coa}) and (\ref{eq:ratio1therm}).

\section{Ratios involving $^4$He}

Based on the idea that a modified $NN$ potential could lead to preclustering effects close to $T_c$~\cite{Shuryak:2018lgd}, more interesting ratios can be considered where the net effect is enhanced~\cite{Shuryak:2019ikv}. A more prominent ratio would involve larger nuclei thus increasing the number of mutual nucleon pairs. In that case one automatically increases the powers of $V/T$ in the exponent of~(\ref{eq:ratio1therm}). However the bigger the nuclei, the larger ``penalty factor'' in the thermal multiplicity~\cite{Braun-Munzinger:2018hat}, as nuclei with increasing mass number $A$ are much more scarce in the final state as can be seen from Eq.~(\ref{eq:yield}).

The nucleus of $^4$He ($=\alpha$), with six mutual distances, provides the optimal case. This state has been measured in low- and high-energy heavy-ion collisions, so it is a good candidate to form light-nuclei multiplicity ratios. The importance of the excited states of $^4$He in the context of thermal models has been addressed in Ref.~\cite{Shuryak:2019ikv}, and put into practice in Refs.~\cite{Lorenz,Vovchenko:2020dmv} for the ${\cal O}_{tpd}$ ratio following the ideas of Refs.~\cite{Hahn:1986mb,Bondorf:1995ua}.

Constructing ratios that cancel most of the factors in Eq.~(\ref{eq:yield}) leads to~\cite{Shuryak:2019ikv}
\be \label{eq:ratio2therm} {\cal O}_{\alpha p^3{\rm He}d} \equiv \frac{N_{\alpha} N_p}{N_{^3\rm{He}} N_d}  \simeq 0.18  \frac{\langle e^{-6V(r)/T}\rangle}{\langle e^{-3V(r)/T} \rangle \langle e^{-V(r)/T} \rangle} \ , \ee
and
\be \label{eq:ratio3therm} {\cal O}_{\alpha tp^3{\rm He}d} \equiv \frac{N_{\alpha} N_t N_p^2}{N_{^3\rm{He}} N_d^3}  \simeq 0.05   \frac{\langle e^{-6V(r)/T}\rangle}{\langle e^{-V(r)/T}  \rangle^3} \ , \ee
where $0.18$ and $0.05$ are the remaining numerical factors after cancellations (the last one corrects the incorrect prefactor quoted in Ref.~\cite{Shuryak:2019ikv} for this ratio).
Notice that in comparison to ${\cal O}_{tpd}$, these ratios are globally enhanced by two and three powers of $V(r)/T$ in the exponential, respectively.

\begin{figure}[ht]
\centering
\resizebox{7.2cm}{!}{\includegraphics{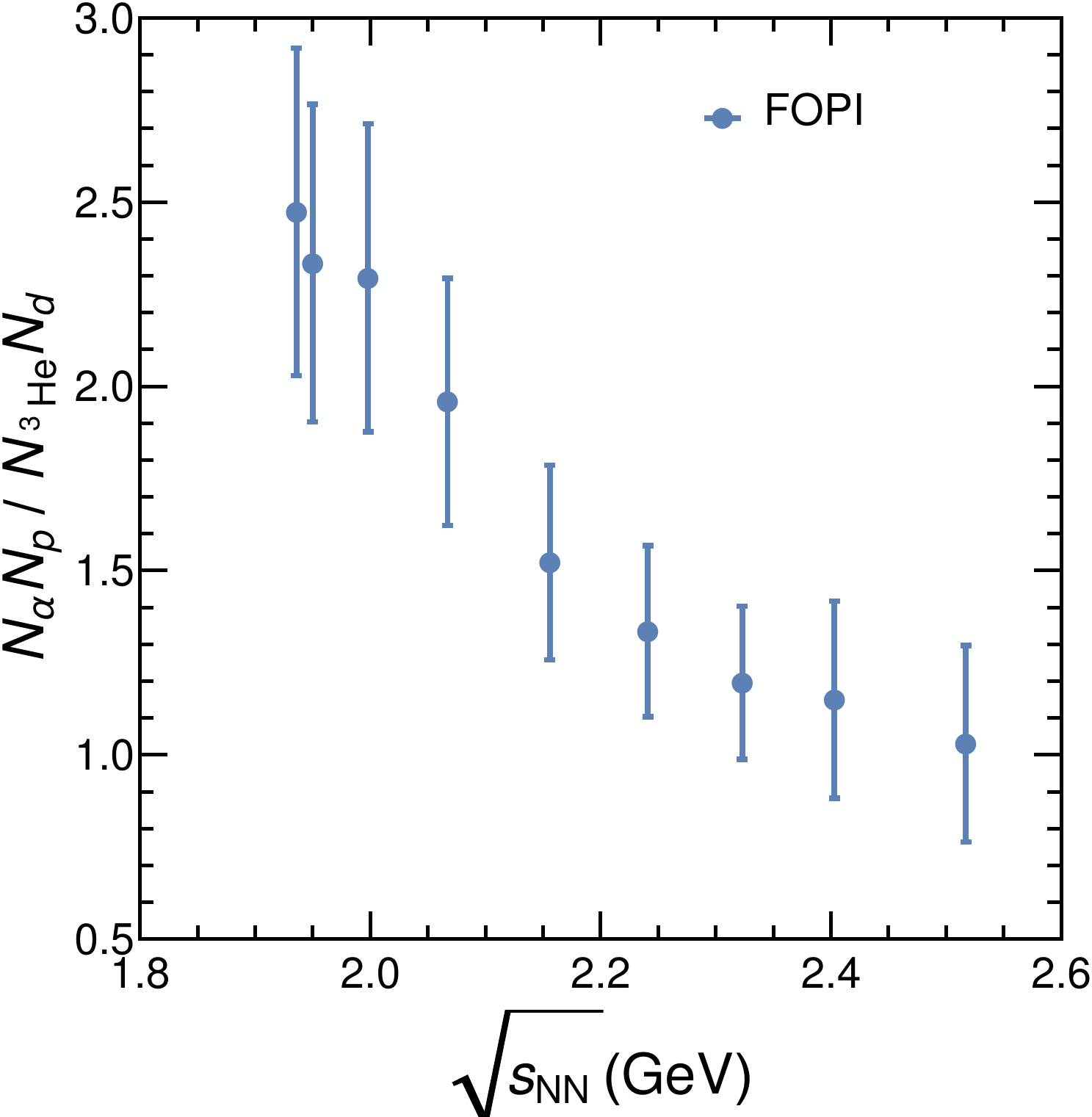}}
\resizebox{7.2cm}{!}{\includegraphics{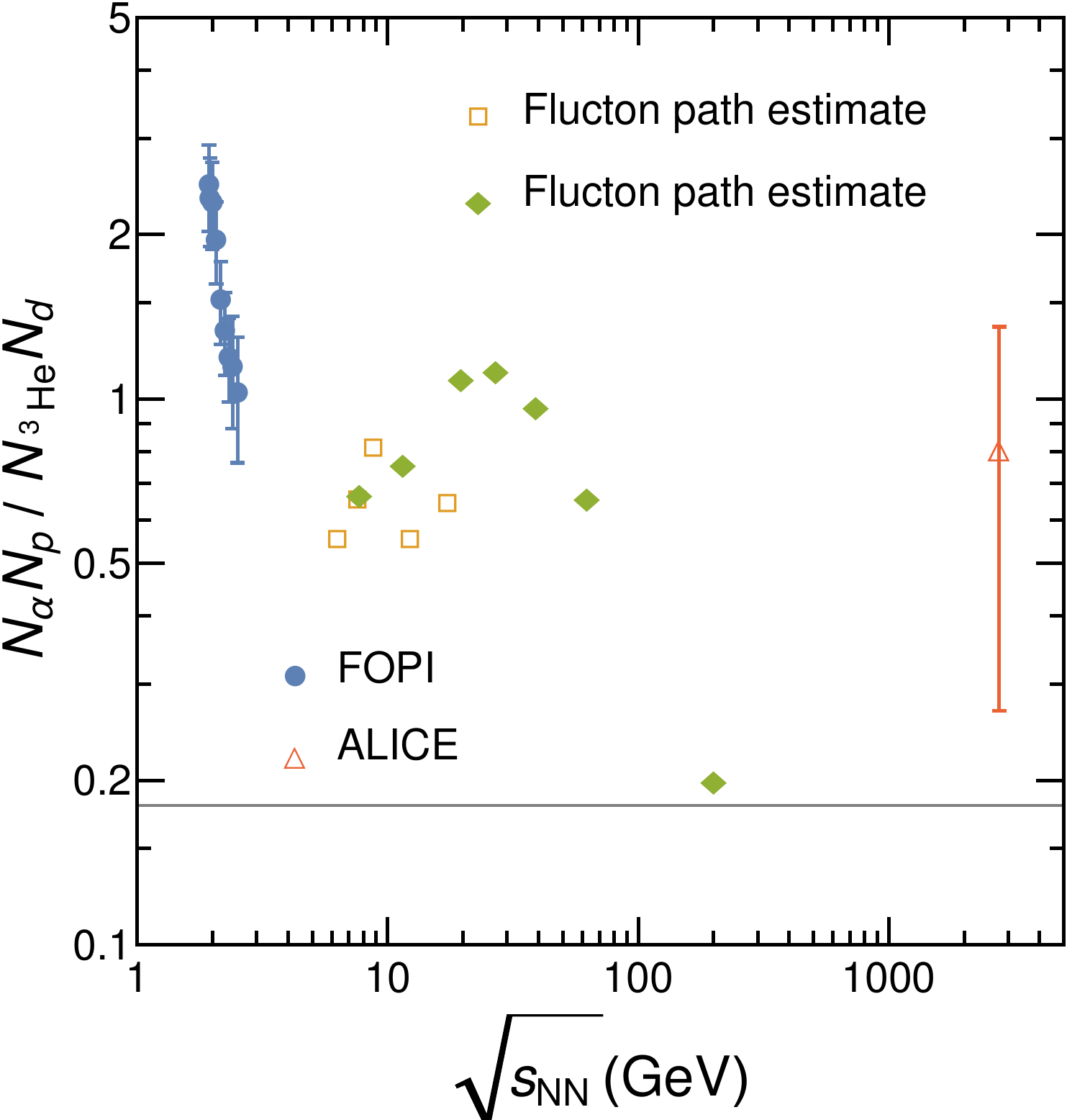}}
\vspace*{0.2cm}
\caption{${\cal O}_{\alpha p^3{\rm He}d}$ ratio as a function of the center-of-mass collision energy. Upper panel: Ratio formed from FOPI multiplicity results~\cite{Reisdorf:2010aa} at low energies. Lower panel: Same FOPI-based data together with ALICE results~\cite{Abelev:2013vea,Adam:2015vda,Acharya:2017bso} at high energies. The ALICE result is formed, not from total multiplicities, but from $dN/dy$ ratios at midrapidity. We provide a theory estimate of such ratio in the intermediate energies (points without error bars) as explained in the text. The horizontal line indicated the value 0.18, cf.~Eq~(\ref{eq:ratio2therm}).}
\label{fig:2}     
\end{figure}

In the upper panel of Fig.~\ref{fig:2} we present the first multiplicity ratio ${\cal O}_{\alpha p^3{\rm He} d}$ built from FOPI yield data~\cite{Reisdorf:2010aa} at low collision energies. As opposed to the previous case, here there is a clear decreasing tendency with $\sqrt{s_{NN}}$. At very low energies might be difficult to make contact with thermal estimates, as the system enters into the multifragmentation region. In the lower panel of Fig.~\ref{fig:2} we add the preliminary result at LHC energies, from ALICE experiment~\cite{Abelev:2013vea,Adam:2015vda,Acharya:2017bso}. As for ${\cal O}_{tdp}$, the top energy FOPI result and the LHC are compatible between them, but should be a coincidence as explained before. Unfortunately, these two sets of experimental data are located at opposed limits in collision energy, away from the expected range corresponding to a possible critical point. We know no experimental data that could be analyzed in that range, calling for an experimental effort to explore that interesting region.

We can offer an estimate of this ratio applying a simple model, and using the NA49 and STAR data for the ratio~(\ref{eq:ratio1therm}) as input. For this we will neglect feed down contributions (which can be added along the lines of~\cite{Vovchenko:2020dmv}), and assume that the structure shown by NA49 and STAR in ${\cal O}_{tpd}$ is entirely due to the effect of a modified potential as described by Eq.~(\ref{eq:ratio1therm}).

We apply the semiclassical thermal flucton method described in~\cite{Shuryak:2019ikv} to the Serot-Walecka potential~\cite{Serot:1984ey} between nucleons.

After a Wick rotation in time, the path integral representation of the density matrix contains the exponential factor of (minus) the Euclidean action of the field configuration, $S_E$. For a pair of nucleons interacting via the potential $V(r)$, the corresponding Euclidean action reads~\cite{Shuryak:2019ikv} 
\be S_E [r(\tau)]=\int_0^\beta d\tau \left[ \frac{m_N}{4} \left( \frac{dr}{d\tau} \right)^2 + V(r) \right] \ . \ee
The periodic configuration that minimizes $S_E[r(\tau)]$, i.e. solution of the classical equation of motion, is the so-called flucton path $r_{{\rm fluc}}(\tau)$~\cite{Shuryak:1987tr}. In the semiclassical approximation we assume that this  flucton path dominates the path integral, so we can compute the probability density of the mutual distances as $P(r)= \exp \{-S_{E} [r =r_{\rm fluc} (\tau)] \}$.
In turn, this probability density allows us to define the spatial average of any observable $A(r)$ by doing
\be \langle A \rangle \equiv \frac{4\pi \int dr r^2 A(r) [P(r)-1]}{4\pi \int dr r^2  [P(r)-1]} \ , \ee
where the $-1$ is used to render the probability density integrable~\cite{Shuryak:2019ikv}, and the denominator normalizes it to one. In this way, the ratios of Eqs.~(\ref{eq:ratio1therm}),(\ref{eq:ratio2therm}),(\ref{eq:ratio3therm}) are directly determined by the flucton solution of the Walecka potential with the $\sigma$ mass and $T$ as parameters. The temperature dependence on $\sqrt{s_{NN}}$ is obtained from the STM as given in Ref.~\cite{Andronic:2017pug}. Therefore, the method utilizes the experimental light-nuclei multiplicity ratios~(\ref{eq:ratio1therm}) from STAR and NA49 experiments to fix the $\sigma$ mass (and therefore set the $NN$ potential) for each collision energy. Then, the same potential is applied to the flucton path method to generate the new multiplicity ratios~(\ref{eq:ratio2therm}) and (\ref{eq:ratio3therm}). This gives an estimate of the expected behavior in the intermediate energy range. Note that the semiclassical method is not appropriate for very low energies where the multifragmentation region and quantum effects start to dominate the nucleon dynamics. In the conditions of high baryon densities the validity of the thermal flucton path is doubtful.

The theoretical points (without error bars) are shown in the lower panel of Fig.~\ref{fig:2}. We observe the same peak structure as in Fig.~\ref{fig:1} but the strength of the effect, i.e. the difference between the top and bottom theory points, is a factor of 2 larger than in the ratio ${\cal O}_{tpd}$.

\begin{figure}[ht]
\centering
\resizebox{7.2cm}{!}{\includegraphics{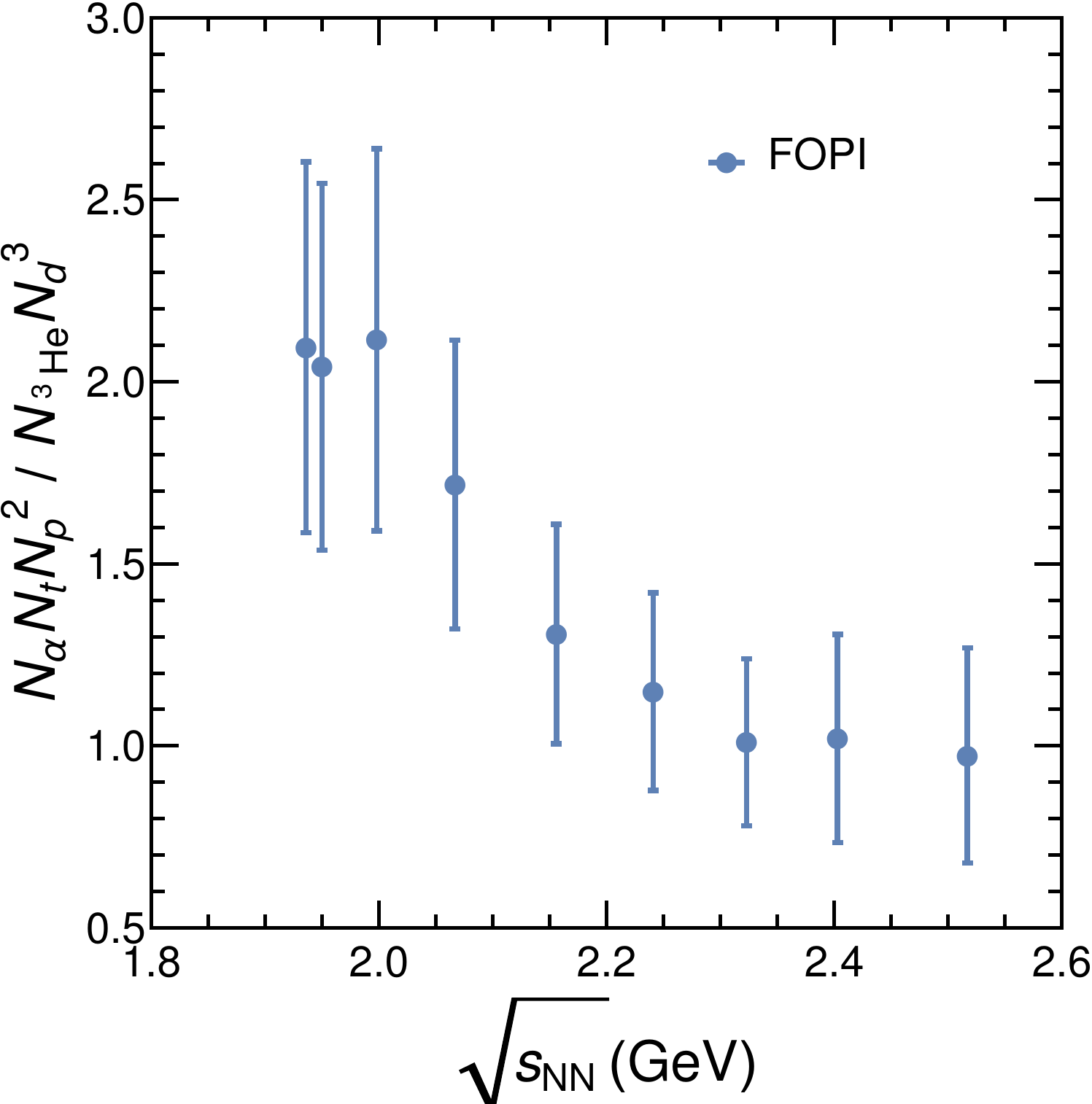}}
\resizebox{7.2cm}{!}{\includegraphics{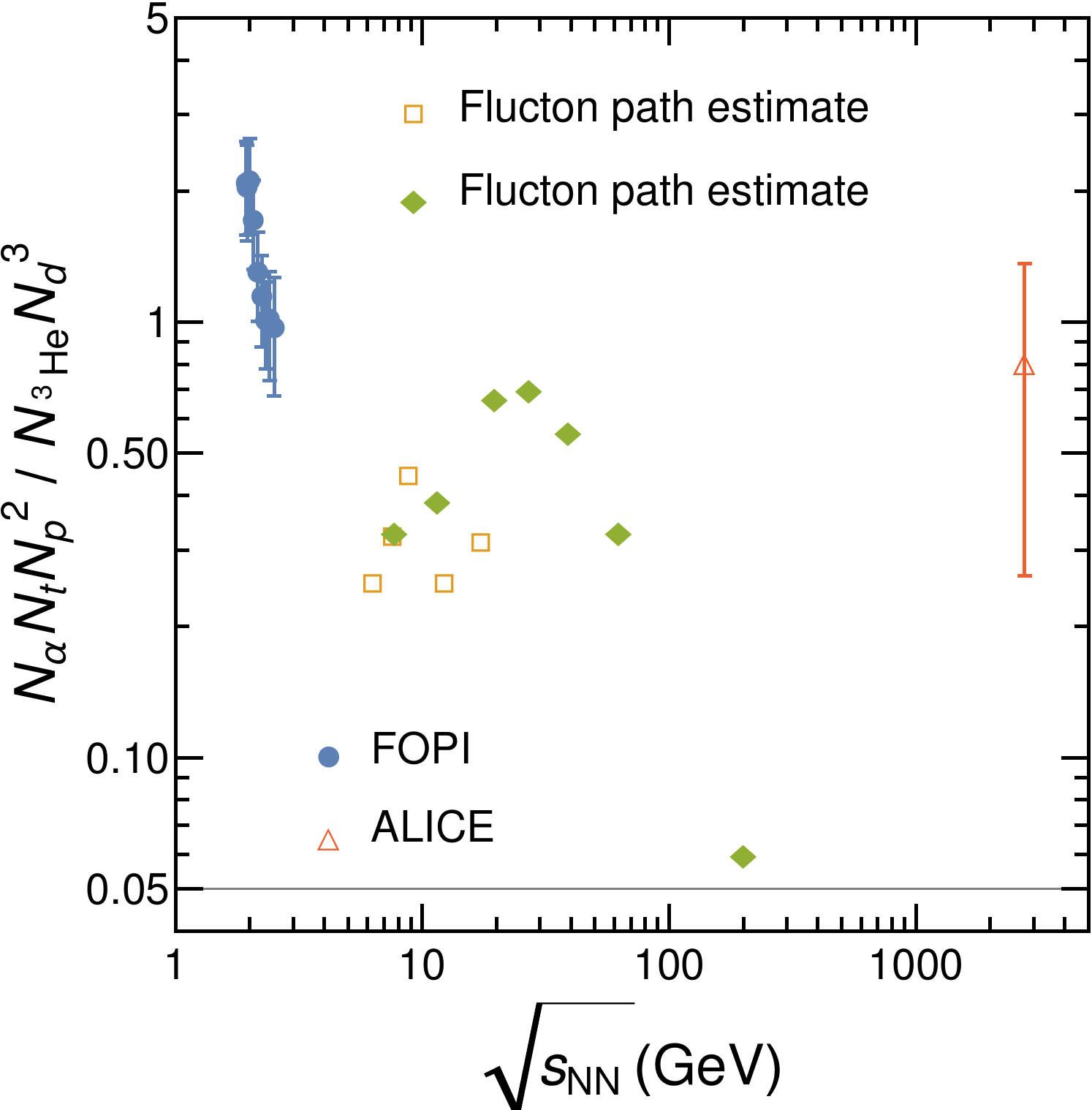}}
\vspace*{0.2cm}
\caption{${\cal O}_{\alpha tp^3{\rm He}d}$ ratio as a function of the center-of-mass collision energy. Upper panel: Ratio formed from FOPI multiplicity results~\cite{Reisdorf:2010aa} at low energies. Lower panel: Same FOPI-based data together with ALICE at high energies~\cite{Abelev:2013vea,Adam:2015vda,Acharya:2017bso}. The ALICE result is formed, not from total multiplicities, but from $dN/dy$ ratios at midrapidity. We provide a theory estimate of such ratio in the intermediate energies (points without error bars) as explained in the text. The horizontal line indicated the value 0.05, cf.~Eq~(\ref{eq:ratio3therm}).}
\label{fig:3}       
\end{figure}

This effect is enhanced in the multiplicity ratio ${\cal O}_{\alpha tp^3 {\rm He} d}$, which we show in Fig.~\ref{fig:3}. The upper panel shows the ratio built with FOPI data~\cite{Reisdorf:2010aa}, which presents again a decreasing pattern with $\sqrt{s_{NN}}$, not seen in ${\cal O}_{tdp}$. Such a pattern deserves more attention in the context of a possible medium-modified nuclear dynamics. In the lower panel we plot ALICE data at $\sqrt{s_{NN}}=2760$ GeV. In this case we again substitute $t$ by $^3$He. Finally, we also plot the theory estimate using the procedure explained before, but applied to the new ratio.

The structure follow the same qualitative pattern as the previous ratio, namely, the top energy FOPI data and LHC are similar, being the intermediate energy prediction lower in magnitude, but with the familiar peak around $\sqrt{s_{NN}}=27$ GeV. Notice that in this case, the strength of the predicted effect (difference between top and bottom points in the intermediate energy region) is a factor of 5 larger with respect to ${\cal O}_{tpd}$, indicating that the signal-to-noise ratio considerably improves in this ratio. This is the reason why we encourage the experimental collaborations to look into these new ratios at intermediate energies.

Finally, we would like to point out that some studies which incorporate partial chemical equilibration, are able to describe light-nuclei production in heavy-ion collisions using kinetic freeze-out temperatures as low as $T_{{\rm kin}} \simeq 100$ MeV~\cite{Xu:2018jff,Vovchenko:2019aoz}. To check the sensitivity to the temperature of our method, we have repeated this calculation for the ratios ${\cal O}_{\alpha p ^3 {\rm He} d}$ and ${\cal O}_{\alpha t p^3 {\rm He}d}$ using a constant $T=100$ MeV for all intermediate collision energies. As this temperature is lower than the temperatures we used from Ref.~\cite{Andronic:2017pug}, one requires less degree of modification of the $V(r)$ (remember that the main effect enters as $|V(r)|/T$). Therefore, we obtain systematically larger $\sigma$ masses, and the final ratios become systematically smaller. The largest differences with respect to Figs.~\ref{fig:2},\ref{fig:3} appear around the peak structures, being in any case no greater than 5\% (the biggest deviation occurs for ${\cal O}_{\alpha t p^3 {\rm He}d}$ at $\sqrt{s_{NN}} = 27$ GeV where using $T=157$ MeV or $T=100$ MeV leads to ${\cal O}_{\alpha t p^3 {\rm He}d}=0.70$ and $0.67$, respectively). Therefore the differences between using $T=100$ MeV and the values of the chemical freeze-out curve in Ref.~\cite{Andronic:2017pug} are insignificant in the flucton method as applied here.

\section{Summary}

In this communication we have proposed new light-nuclei multiplicity ratios (\ref{eq:ratio2therm})(\ref{eq:ratio3therm}) involving $^4$He ($=\alpha$), 
\be {\cal O}_{\alpha p ^3 {\rm He} d} \ , \qquad {\cal O}_{\alpha t p^3 {\rm He}d} \,
\ee
which are sensitive to the nonideal production of light-nuclei in heavy-ion collisions. Should the QCD critical point enhance the attraction between nucleons to the extend of an increase of light-nuclei formation, then these ratios would present maxima as functions of the collision energy. Such peaks would be more compelling than the one considered so far in Eq.~(\ref{eq:ratio1coa}), as the variability in ${\cal O}_{\alpha p ^3 {\rm He} d}$ and ${\cal O}_{\alpha t p^3 {\rm He}d}$ is, respectively, a factor of 2 and 5, compared to ${\cal O}_{tpd}$.

We have demonstrated the feasibility of such measurements by presenting experimentally-based multiplicity ratios at low and high energies by FOPI and ALICE experiments, respectively. We have also motivated the usefulness of our proposal by computing a theoretical estimate of these new ratios at intermediate energies based on previous NA49 and STAR data for the ratio of Eq.~(\ref{eq:ratio1therm}). We exploited the semiclassical thermal flucton method applied to the Serot-Walecka potential, as detailed in Ref.~\cite{Shuryak:2019ikv}. 

We hope that experiments running at intermediate energies can address these observables in the future to assess the possible effect of critical dynamics via light-nuclei production.

\begin{acknowledgement}
We acknowledge Benjamin D\"onigus for bringing the FOPI and LHC data on light-nuclei production to our attention, and for useful discussions. 

Work supported by the Office of Science, U.S. Department of Energy under Contract No. DE-FG-88ER40388 and by the Deutsche Forschungsgemeinschaft (DFG, German Research Foundation) through Projects No. 411563442 (Hot Heavy Mesons) and No. 315477589 - TRR 211 (Strong-interaction matter under extreme conditions).
\end{acknowledgement}

\end{document}